\documentstyle[12pt]{article}

\begin{document}
\sloppy
\thispagestyle{empty}

\setcounter{page}{0}

\vspace{-1cm}

Interner Bericht      \\
DESY--Zeuthen 95--06  \\
Dezember 1995         \\

\mbox{}
\vspace*{\fill}
\begin{center}

{\LARGE\bf  Measurement of Singly Polarized}\\

\vspace{3mm}
{\LARGE {\bf $p\vec{N}$ Collisions at HERA}
    \footnote{Contribution to the Proceedings of the {\it Workshop
    on the Prospects of Spin Physics at HERA}, held at Zeuthen
    August 28-31, 1995}}  \\

\vspace{2em}
{\large
Wolf-Dieter Nowak}
 \\
\vspace{2em}
{\large
{\it  DESY--IfH Zeuthen,}
 \\
{\it 15735 Zeuthen,   Germany}}     \\ [15pt]
{e--mail: nowakw@ifh.de}            \\
\end{center}
\vspace*{\fill}
\begin{abstract}
\noindent
A summary is given on  the physics potential of measuring singly polarized
proton--nucleon collisions using a polarized internal target in the
820 GeV HERA proton beam.
This summary is based upon talks given at the 2nd Meeting on
{\it Possible Measurements of Singly Polarized $p\vec{p}$ and $p\vec{n}$
Collisions at HERA},
which was held at DESY Zeuthen from August 31 to September 2, 1995
as a follow-up to the {\it Workshop on the Prospects of Spin Physics at HERA}.
\end{abstract}
\vspace*{\fill}
\newpage

%
\section{Introduction}
\label{sect1}

A series of meetings was started at PNPI Gatchina in July~1994 \cite{gat}
to study the physics case for possible measurements of singly polarized
proton-proton and proton-neutron collisions utilizing an internal polarized
H/D target in the unpolarized 820~GeV HERA proton beam. Such a target offers
unique features as polarization above 80\%, no dilution, and a high density
up to $10^{14}$ atoms/cm$^2$ \cite{ste1}. Moreover, small systematic errors
are expected when comparing data from protons and neutrons.  \\

Measurements of single spin asymmetries were performed a few years ago by
the E704 collaboration at Fermilab \cite{yok1,nur1}. Using an unpolarized
Hydrogen target in an external 200~GeV polarized proton beam intriguingly large
single transverse spin asymmetries $A_N$ were revealed in inclusive
pion production \cite{704a}. At transverse momenta between 0.7 and 2~GeV/c
$A_N$ exhibits a clear increase with $x_F$, as can be seen from fig.~1.
Similar results had been obtained earlier at BNL \cite{BNL1} and IHEP
\cite{IHEP} utilizing beam energies of 18 and 40 GeV, respectively.   \\
In the framework of perturbative QCD inclusive
aymmetries $A_N$ are supposed to vanish at twist-2 and significant non-zero
asymmetries are usually associated with higher twist effects. However, the
experimental asymmetries are much larger than those predicted by
pQCD. To this end, the data seem to seriously question the validity of the
presently accepted pQCD picture, at least the onset of its validity is not
yet reached at the presently accessible transverse momenta.   \\

Measuring single spin asymmetries at HERA with $\sqrt{s} \simeq
40$~GeV would open up the possibility to extend the reach in transverse
momentum up to about~10~GeV/c depending on the inclusive final
state chosen. This appears feasible in three real years of HERA running
allowing for an integrated luminosity of about 240 pb$^{-1}$ \cite{now1}.
Data taken in this unexplored kinematic region would
yield very interesting input to further understand the transition region
between non-perturbative and perturbative QCD and thus facilitate a deeper
understanding of the QCD spin sector \cite{bro1}.  \\
If such an experiment utilizing a polarized target in the HERA proton beam
(tentative project name HERA--$\vec{N}$) should
become feasible in a few years, the unpolarized beam would focus its initial
physics scope to the measurement of single spin asymmetries ('phase~I').
Once later polarized protons should become available, the same set-up would be
readily available to measure various kinds of double spin asymmetries.
These 'phase~II' measurements would constitute an alternative --fixed target--
approach to similar
physics as it will be accessible to the STAR and PHENIX experiments
at the low end of the RHIC energy scale ($\sqrt{s}~\simeq~50$~GeV) \cite{bun1}.
The two
most interesting spin physics items at this time will presumably still be the
separate measurements of the polarized gluon and strange quark distributions.
Their knowledge is badly needed to eventually resolve the long--standing
problem of the nucleon spin decomposition.   \\

The aim of this meeting was to present theoretical results obtained over
the past 12 months on single spin asymmetries in various inclusive final
states. Future directions of interest were subject of a lively discussion
led by J. Soffer. A
summary of this meeting, selected from the viewpoint of
the feasibility to perform adequate measurements in phase~I of
HERA--$\vec{N}$,  will be given in the subsequent
chapters of this paper. Two contributions dealing with 'phase II' physics
\cite{tan1,pir1}
will not be discussed here. The full text of all contributions can be found
in a separate proceedings volume \cite{Proc}.   \\
The 3rd meeting of the series is scheduled to be
held at JINR Dubna in June~1996, its physics scope will be extended
to cover both phase~I and phase~II of HERA--$\vec{N}$ physics.

\vspace{0.5cm}

%
\section{Review Talks}
\label{sect2}

In the invited introductory talk S. Nurushev \cite{nur2} discussed
the variety of existing data on single spin asymmetries measured at beam
energies above 10 GeV. In the detailed review on single spin measurements
performed with the E704 experiment \cite{nur1} the unexpectedly large $A_N$
measured at average transverse momenta of about 1~GeV/c
appears still as one of the most interesting
pieces of data which calls for confirmation at higher energies.  \\

An invited survey on measurements of single spin asymmetries at RHIC was
presented by G. Ladinsky \cite{lad1}. At RHIC both beams will be polarized from
the very beginning, hence double and single spin asymmetries will be accessible
at the same time. Data taking is envisaged to yield 800 inverse pb$^{-1}$ at
$\sqrt{s}$~=~500~GeV and 320 pb$^{-1}$ at $\sqrt{s}$~=~200~GeV.   \\
At these energies and luminosities W$^{\pm}$ and Z physics can be studied and
parity violation in electroweak interactions can be used as a tool. In
particular, the parity violating helicity asymmetry $A_L^{PV}$ allows to study
the polarized antiquark distributions $\Delta \bar{u}$ and $\Delta \bar{d}$
with
accuracies on the few percent level. Also, $A_L$ measurements in Drell--Yan
pair production can be used to probe sea quark densities.
The measurement of single transverse spin asymmetries in inclusive photon
and pion production at RHIC might provide information on twist-3 parton
distribution functions, although they are presently expected to be very small
at the anticipated transverse momenta of several 10 GeV.  \\

The today's knowledge on the spectrum of single spin asymmetries potentially
accessible at HERA was
covered in two mini-review talks by G. Ladinsky \cite{lad2} and O. Teryaev
\cite{ter1}. Most of the topics included there will be discussed in the
following sections, although from an experimentalist's point of view.

\vspace{0.5cm}
%
\section{QCD Based Single Spin Asymmetries}
\label{sect3}

{\bf Photon Asymmetries.}   \\
Quantitative estimates of single transverse spin asymmetries measured in
inclusive direct photon production were discussed by A. Sch\"afer \cite{sch1}.
Sizeable asymmetries would indicate the existence of (leading) twist-3 effects
and thus allow to study the almost virgin field of multi--parton correlations
in the nucleon. In contrast to earlier
predictions of J. Qiu and G. Sterman \cite{qiu1}
suggesting an asymmetry of 10...20\%  more recent estimates of quark--gluon
correlations by A. Sch\"afer et al. \cite{sch2} indicate asymmetries of only
a few percent when measuring with a proton target ; the
neutron target asymmetry is predicted to be several times larger.  \\
Basing on the possible existence of a collective gluon field in the nucleon
a direct relation was revealed some time ago \cite{sch3}
between these asymmetries and $d^{(2)}$, the
twist-3 contribution to the second moment of the deep inelastic structure
function $g_2$, and predictions for $d^{(2)}$ were calculated. Predictions were
obtained by other groups
from QCD sum rules and Lattice calculations, however up to now
there is no clear consensus between them \cite{man1}.  \\
On the other hand, recent preliminary data from E143 \cite{roc1}, although
measured at rather small $Q^2$,  indicate very
small values for both $d^{(2)}_p$
and $d^{(2)}_d \simeq \frac{1}{2} (d^{(2)}_p + d^{(2)}_n)$. Moreover, within
error bars no difference is seen yet between results from proton and neutron
target, respectively.
Hence more precise data on $g_2$ are required to be able to decide whether the
measurement of single transverse spin asymmetries in inclusive photon
production with HERA--$\vec{N}$ could
be a useful tool to study the possible existence of a coherent gluon field in
the nucleon.  \\

The contribution to the same asymmetry basing upon 3--gluon correlations
\cite{xji1} instead of quark--gluon ones was calculated, as well \cite{sch1}.
The resulting hard scattering asymmetry increases with
$\sqrt{s}$, however at HERA--$\vec{N}$ energies it does barely exceed the 1\%
level, as can be seen from fig.~2. It might therefore be very difficult to
untangle this contribution.   \\

{\bf Jet Asymmetry.}   \\
In a dedicated talk given at the main workshop O. Teryaev discussed
twist-3 aspects in proton--nucleon single spin asymmetries \cite{ter3}.
He recalled that the contribution from 'gluonic poles', introduced by J. Qiu
and G. Sterman and reconsidered by A. Sch\"afer et al., is now believed to
be small  (cf. previous section). However the 'fermionic poles', suggested
even earlier by A. Efremov and O. Teryaev as a possible source of single spin
asymmetries \cite{efr1}, might in fact yield the dominant contribution.   \\
Based upon the hypothesis of  'fermionic poles' the short--distance part was
already calculated in Born approximation for a number of quark--gluon
subprocesses \cite{ter4}. Some properties of the large distance partonic
correlations were obtained by means of a simple model for the
twist-3 part of $g_2$ that accomodates the sum rules derived in twist-3 QCD.
It is compatible with existing data but smaller error bars are required to
eventually decide on its validity.   \\

In case of jet production the simple left/right cross section asymmetry
(with respect to the normal to the scattering plane) is found to represent
the asymmetries of quark and gluon production and claimed to be as clear a
QCD test as the direct photon asymmetry. Fragmentation is not involved
at this level, hence no complicated detailed analysis of the jet structure is
required. Whereas the short--range calculations support a jet asymmetry
on the 10\% level the long--range estimates based on the model of
twist-3 $g_2$ and QCD
sum rules indicate an asymmetry on the level of a few percent, only.
The long--range part needs to be understood better before
quantitative predictions including the $x_F$ dependence of the asymmetry
can be given. \\
Under HERA--$\vec{N}$ conditions the high jet cross section would allow for
statistical errors on the permille level \cite{lad2,now1}. Hence the
experimental sensitivity will be dominated by the systematic error which has
to be assessed carefully.  \\

It was indicated that very similar theoretical arguments hold for a left/right
single transverse spin asymmetry in dilepton production \cite{ter3}. No
quantitative predictions exist here either.  \\

{\bf Pion Asymmetries.}   \\
Since the quark--gluon subprocesses contribute also to single spin asymmetries
in pion production a possible dominance of 'fermionic poles' over 'gluonic
poles'
would lead to very interesting conclusions. For practically vanishing gluonic
poles significant numerical differences are observed if instead of a gluon a
quark in the transversely polarized nucleon is struck by the gluon of the
unpolarized nucleon \cite{ter3}; the asymmetry due to the gluon--quark
contribution is obtained several times larger \cite{ter5}. Even more
importantly, this mechanism quite 'naturally' leads to a flavour dependence
when explaining the pion
asymmetries; the $\pi^+$ asymmetry is related to the correlation between
gluon and $u$--quark, the $\pi^-$ asymmetry to the one between gluon and
$d$--quark.   \\
No quantitative predictions were presented yet, but the existence of mirror
asymmetries for $A_{\pi^+}$ and $A_{\pi^-}$ can already be deduced rather
easily.   \\

{\bf Dimuon Asymmetry.}   \\
As for any inclusive final state, single helicity asymmetries vanish on the
Born level for Drell--Yan production, as well. At order $\alpha_s$ a non--zero
asymmetry can be constructed by the interference of tree diagrams with
1--loop diagrams, provided the dilepton pair has a non--vanishing transverse
momentum. Hence both momenta of the pair have to be detected to access the
asymmetry.   \\
Following the initial paper of R. Carlitz and R. Willey \cite{car1} P. Nadolsky
has calculated predictions for HERA--$\vec{N}$ conditions \cite{nad1}. There
the steeply falling cross section, suppressed by $\alpha_{em}$ as in
case of inclusive direct photon production, requires to consider rather small
dilepton masses Q$^2 \simeq$ 1.5 (GeV/c)$^2$
and similar transverse momenta Q$^2_{\perp}$ at the same
time to end up with a reasonable statistics. Special care had to be taken to
alleviate several theoretical difficulties arising in this kinematic region.
Still, as shown in fig.'s~3a and 3b for Q$^2$~=~1.3 and 2~(GeV/c)$^2$,
even there the eventually achievable statistics is not sufficient to
distinguish between $\Delta$G~=~0 and $\Delta$G~=~2; only very large values
like $\Delta$G~=~5 would lead to statistically significant effects.   \\
As a remaining possibility the 'dimuon + jet' final state might have better
sensitivity, since there the parton cross section needs not to be
integrated over x$_{Bj}$. A corresponding study is in progress \cite{nad2}.  \\

{\bf J/$\psi$ Asymmetry.}   \\
A novel way to possibly access the gluon contribution to the nucleon spin
was discussed by O. Teryaev \cite{ter2}. Inclusive helicity asymmetries are
supposed to vanish on the Born level in singly polarized nucleon--nucleon
collisions. Preliminary results from one-loop calculations were shown for
$\sqrt{s}$~=~40 GeV. As can be seen from fig.~4a, a hard scattering asymmetry
of a few percent is expected for not too large J/$\psi$ energies at
$p_t^2$~=~5 (GeV/c)$^2$, which is about the lower limit of the expected
HERA--$\vec{N}$ acceptance. The calculated cross section, shown in fig.~4b,
appears high enough to realize statistical asymmetry errors on the permille
level \cite{now1}. This might be sufficient since the only dilution in
J/$\psi$ production comes from target polarization times average initial gluon
polarization, which is anticipated to be of the order of 0.2 to 0.4. Certainly
Monte Carlo studies and a careful assessment of the systematic errors
are required to decide on the feasibility of this very
interesting measurement.   \\

{\bf $\Lambda$--Hyperon Asymmetry.}   \\
The measurement of inclusive $\overline{\Lambda}^0$ production in singly
polarized proton--nucluon scattering implies the possibility to determine
the strange quark contribution to the nucleon spin, $\Delta s$. However,
in the polarized target nucleon the hyperon can be produced from both
a strange quark or a gluon at similar x$_{Bj}$. This makes it difficult
to determine $\Delta s$ separately from the gluon spin contribution,
$\Delta G$.   \\
In a talk by S. Manayenkov \cite{man1} it was outlined that indeed it is
practically impossible to untangle both contributions from the analysis of
kinematical variables. Another way to probe $\Delta s$ might be to measure the
$\overline{\Lambda}^0$ polarization. Under HERA--$\vec{N}$ conditions it
was found from Monte Carlo calculations that this polarization can be
determined for transverse momenta up to 5 GeV/c. There are indications that
the statistical accuracy will allow to distinguish between different strange
quark polarizations.

\vspace{0.5cm}
%
\section{Models with Orbital Angular Momentum}
\label{sect4}

Trying to explain the non-zero transverse pion asymmetries measured by
E704 some theoretical models are successful when assuming the existence
of quark orbital momenta in the polarized nucleon. Two of them were discussed
during the meeting.  \\

{\bf Current Quark Orbital Momentum in the Constituent Quark.}   \\
In the constituent quark model approach of S. M. Troshin and N. E. Tyurin
\cite{tro1}, presented by S. M. Troshin \cite{tro2}, the current quarks are
assumed to perform an angular motion 'inside' the constituent quarks.
The latter
are believed to have a relatively slow motion within the whole nucleon. The
origin of the current quark orbital angular momentum is explained by pairing
correlations in analogy with an anisotropic generalization of the theory of
superconductivity \cite{fri1}.  An axis of anisotropy is
associated with the polarization vector of the valence quark, located at the
origin of the constituent quark. A non-zero orbital momentum of the partons
inside the constituent quark implies their existence of significant multiparton
correlations, hence this picture is well suited to describe a hadron at
moderate momentum transfers.
At high momentum transfers the internal structure of a constituent quark is
resolved and it is represented as a cluster of non--interacting current
quarks.   \\

{\it Pion Asymmetries.}   \\
Compensation effects between valence and sea quark spins can lead to a $p_t$
dependent asymmetry for the inclusive meson which is formed by
constituent quark recombination. The resulting
single spin asymmetry shows a weak energy dependence, significant values start
at transverse momenta above 1~GeV/c \cite{tro1}.
The predicted $A_N$ values can be seen from fig.~5a,
where inclusive $\pi^0$ production at laboratory momenta of 200~GeV and 800~GeV
is shown together with E704 data. Corresponding predictions for inclusive
charged pion production are shown in fig.~5b.   \\

{\it $\phi$ Asymmetry.}   \\
Applying the same model to inclusive $\phi$ production \cite{tro3}
the magnitude of the non--zero orbital momentum of rotating strange
quark--antiquark pairs is found proportional to the magnitude of the
polarization of the constituent, i.e. the valence quark. The proportionality
factor is just given by the relative amount of strange quarks within the
constituent quark. At moderate $p_t~\simeq$~1~GeV/c the asymmetric structure
of the constituent quark is probed and leads to a single spin asymmetry
predicted to be of the order of 1 to 5\%.   \\

{\bf Valence Quark Orbital Momentum in the Nucleon.} \\
In the 'Berliner Model' of T. Meng and co-workers \cite{meng}, presented by
Z. Liang \cite{lian}, the orbital angular momentum is assigned to the valence
quarks directly. Since orbital momentum is not a good quantum number, orbital
motion is involved for quarks even when being in the ground state. From the
semi-classical wave function for a polarized nucleon it is deduced
that the average number of valence quarks with a polarization parallel
to the nucleon polarization is different from the number of those being
polarized anti-parallel. Hence this relativistic quark model avoids from the
very beginning the 'deficiency' of perturbative QCD of not being flavour
sensitive.  \\

{\it Pion Asymmetries.}   \\
A significant part of mesons observed in the fragmentation region of the
polarized nucleon are believed to be directly
created from a valence quark and a sea anti-quark. The effective orbital motion
attributed to the valence quark in conjunction with a significant (front)
surface effect encountered by the unpolarized nucleon leads then, in the
fragmentation region of the polarized nucleon, to significant single spin
asymmetries whose magnitude increases with rising $x_F$. It has to be
mentioned here that the existence of a surface effect of the necessary strength
is not unanimously agreed upon. Still, there is fair
agreement with the E704 inclusive pion data, as can be seen from fig.~6.
The predictions are not sensitive to the type of the incoming unpolarized
nucleon and no asymmetry is expected in its fragmentation region.     \\

{\it Drell--Yan and W/Z Production.}   \\
Predictions of the same model, applied to calculate left-right asymmetries in
singly polarized Drell-Yan pair and W/Z-boson production processes, were
presented by C. Boros \cite{boro}. For Drell--Yan pairs produced by two
nucleons the
above described creation process leads to non-zero asymmetries not only in the
fragmentation region of the polarized nucleon, but also in the region of small
positive and negative $x_F$-values. As can be seen from fig.~7, given
for 820~GeV incoming energy, the flavour
sensitivity implied in the model leads to substantially different predictions
for polarized protons, neutrons, or deuterons.   \\
Additionally, predictions have been calculated for Drell--Yan pair production
by pions as well as for W/Z-boson production in singly polarized
(anti) nucleon--nucleon collisions. Strongly different asymmetries are
predicted at $\sqrt {s}$~=~500~GeV
for $W^+$ vs. $W^-$ production which could be tested at RHIC or possibly at
the Tevatron with a polarized proton beam.  \\

\vspace{0.5cm}
%
\section{Other Models}
\label{sect5}

{\bf Double J/$\psi$ Production in the Colour Singlet Model.}   \\
The production of J/$\psi$ pairs was considered by S. Baranov and
H. Jung \cite{bar1} as a tool to probe the gluon polarization using the
framework of the non--relativistic Colour Singlet Model \cite{hum1}.
Recently, T. Gehrmann reconsidered the case for HERA--$\vec{N}$ \cite{ger1}.
At $\sqrt{s} \simeq 40$~GeV gluon--gluon hard scattering dominates over
quark--antiquark annihilation, hence the majority of double J/$\psi$ events
could be used to probe the gluon polarization. However, the cross section for
p$_t^2~\geq$~1~(GeV/c)$^2$ amounts to about 5 pb, only. This is further
diminished by a branching fraction of at most 12\% in conjunction with an
about 20\% chance to reconstruct the helicity state of one of the two
J/$\psi$'s. This results in a combined efficiency of about 2\%, even without
taking into account the experimental acceptance. Altogether this
yields about 30 useful events expected during the three (assumed) years of
HERA--$\vec{N}$ running, i.e. there is no realistic chance to access the
gluon polarization through single spin asymmetry measurements in double
J/$\psi$ production.  \\

{\bf Instanton Mechanism of Single Spin Asymmetries.}   \\
Non--perturbative interactions between quarks can be associated with the
existence of instantons, i.e. vacuum fluctuations of gluon fields.
N. Kochelev \cite{koc1} obtained the instanton contribution to the quark
distribution functions from the cross section for the instanton induced
quark--nucleon interaction. The latter is found to be characterized by both a
spin--flip and a strong flavour dependence, which makes it an especially
interesting
mechanism when considering flavour dependent effects in polarized scattering,
like e.g. single spin asymmetries as seen for inclusive pions.   \\

Whereas single spin asymmetries haven't been considered quantitatively, yet,
predictions were shown for the total quark contribution to the nucleon spin,
$\Delta \Sigma$, and the polarized longitudinal structure function $g_1(x)$.
The finite size of the instantons, described by a form factor, results in a
strong Q$^2$--dependence of $\Delta \Sigma$. As can be seen from fig.~8,
the violation of the Ellis--Jaffe sum rule appears in the instanton model
at low photon virtualities below 2 (GeV/c)$^2$, i.e. is clearly a
non--perturbative effect. Similarly, the predictions for the polarized
structure functions $g_1^p(x)$ and $g_1^n(x)$, shown in fig.'s~9a and 9b,
exhibit instanton induced effects at lower values of x$_{Bj}$, only. These
predictions can be clearly checked against the
forthcoming precision data from the HERMES, E154, and E155 experiments.   \\

{\bf Diffractive High p$_t$ 2--Jet Production.}   \\
Another possibility to study the spin structure of QCD at large distances,
i.e. the hadron wave function, may be given by the p$_t$--dependence of
the single transverse spin asymmetry in diffractive $Q \bar{Q}$ production.
In a perturbative calculation of the spin--dependent quark--pomeron vertex
\cite{gol1} additional contributions were identified which imply significant
spin--flip terms. The latter were shown not to affect the standard pomeron
contribution to the proton structure function.   \\
Using this spin--dependent description the single spin asymmetry for
diffractive light quark $Q \bar{Q}$ production was calculated at the
pomeron--proton vertex and compared to the standard case \cite{gol2}. Results
obtained for $\sqrt{s} \simeq 40$~GeV at small pomeron fractional momentum
(x$_p$~=~0.05) and small momentum transfer ($\mid$ t $\mid$~=~1~GeV$^2$) are
shown in fig.~10. Additionally, the experimental sensitivity is shown by the
projected error bars for HERA--$\vec{N}$ conditions without taking into
account the acceptance of the
spectrometer. As can be seen, the standard pomeron implies no p$_t$--dependence
for the asymmetry, whereas by the spin--flip contributions a
remarkable transverse momentum dependence is predicted which has a good chance
to be confirmed after  an integrated luminosity of 240 pb$^{-1}$ has been
collected.  \\
It is important to note that besides the two high p$_t$ jets it is necessary to
detect the recoil hadron to guarantee the diffractive nature of the collision.

\vspace{0.5cm}
%
\section{Conclusions}
\label{sect6}

As it is well known, single spin asymmetries for inclusive final states
are supposed to vanish in the
framework of perturbative QCD when considering lowest order for helicity
asymmetries and twist-2 in the case of transverse spin, respectively.
Hence the intriguingly large single transverse spin asymmetries measured
at BNL, IHEP, and by E704 for 0.5 GeV$^2~\leq$~p$_t^2~\leq$~4~GeV$^2$
are preferentially to be interpreted as higher
twist effects. This hypothesis would be confirmed if an experiment accessing
larger transverse momenta (like HERA--$\vec{N}$) should find these
asymmetries decreasing with p$_t$, as predicted by pQCD.
As a result the onset of pQCD would be defined more exactly and, more
generally, the pQCD spin sector would be validated.  \\

On the other hand, it is quite possible that the envisaged exploratory
measurements with HERA--$\vec{N}$ (phase I) might reveal the persistence
of large asymmetries at higher transverse momenta. There are indeed
strong indications from several sides that higher twist effects
might not vanish at all at rather large p$_t^2$, in strong contrast to pQCD
expectations. Very recent theoretical results on the proton magnetic form
factor, obtained from leading twist calculations \cite{bol1}, fall short by at
least 50\% when compared to the data from unpolarized exclusive reactions.
The striking non--zero single spin asymmetry found more than 10 years ago
at the AGS in 28 GeV/c proton--proton elastic scattering at
p$_t^2~\simeq$~6~GeV$^2$ \cite{cam1} is at variance with pQCD since that
time.   \\
These and other disagreements between theory and experiment
might imply that the simple twist-2 picture of perturbative QCD
is not applicable, at least to certain reactions. If this was true, the
envisaged measurements with HERA--$\vec{N}$ with their adequate reach in
p$_t$ would allow to shed completely new light onto the structure of
Quantum Chromo Dynamics at
small distances. It is not difficult to predict that in this case the
theoretical interest to perform such an experiment would presumably be
even higher.   \\

In any case, new data about the p$_t^2$ dependence of single spin asymmetries
in an hitherto unexplored region up to several tens of GeV$^2$ should allow
for a considerable improvement in the understanding of the features of
quark--gluon interactions. Especially, such
data might turn out to be very useful to be confronted with then much
further developed QCD lattice calculations.   \\

In summary, the 2nd Meeting on {\it Possible Measurements of Singly Polarized
$p\vec{p}$ and $p\vec{n}$
Collisions at HERA} has demonstrated a considerable improvement
in understanding the physics potential of a possible HERA--$\vec{N}$
experiment. Still, some more general and some additional specific
components would make the physics program more sound. We need
quantitative predictions for single transverse spin asymmetries
based upon the proposed existence of 'fermionic poles'. Also, calculations
from non--perturbative models, like e.g. the instanton model, might lead to
very interesting and possibly unexpectedly large predictions on single spin
asymmetries in the HERA--$\vec{N}$ domain, i.e. for transverse momenta up to
10~GeV. A realistic way to access $\Delta$G via single spin asymmetries
would be a very desirable extension of the physics scope.
Generally speaking, more plausible and compelling theoretical ideas
are most welcome to further increase the interest in such an experiment.  \\

%
\section*{Acknowledgements}
\label{sectn}

The 2nd Meeting on {\it Possible Measurements of Singly Polarized
$p\vec{p}$ and $p\vec{n}$ Collisions at HERA}
would not have been possible without the effort of many
theoreticians fascinated on one hand about the rich physics associated
with single spin asymmetries and interested in setting up a viable physics
program for an experiment at HERA which might measure them, on the other.   \\
Special thanks goes to Jacques Soffer for his permanent advice on spin
physics and his effort in moderating many hours of lively debates.
Numerous discussions with Glenn Ladinsky and Oleg Teryaev, especially in
Zeuthen during the weeks preceding the workshop, are warmly acknowledged. Also
highly recognized is the care of Bernard Pire about the subject of the meeting
and the enthusiasm of Thomas Gehrmann and Pavel Nadolsky devoted to the
investigation of specially interesting channels. Many thanks to Stan Brodsky,
Nikolai Kochelev, Peter Kroll, and Andreas Sch\"afer
for very stimulating discussions, and to Helmut B\"ottcher for
carefully checking of the manuscript.


\newpage
%
\section*{Figure Captions}
\label{sectn1}

\begin{itemize}
\item[Fig.~1:]
         Single transverse spin asymmetry $A_N$ in dependence on $x_F$,
         as measured by E704 for inclusive pion production at
         small and medium transverse momentum.
\item[Fig.~2:]
        Single transverse spin asymmetry vs. $x_F$, as predicted from twist--3
        pQCD 3--gluon correlations
        for inclusive direct photon production at $\sqrt{s}$ = 30 GeV.
\item[Fig.~3:]
        Single helicity asymmetry for Drell--Yan production vs. $Q_{\perp}^2$,
        as predicted from pQCD one--loop calculations for
        $Q^2$~=~1.3~(Gev/c)$^2$ (fig.~3a) and
        $Q^2$~=~2~(Gev/c)$^2$ (fig.~3b) at $\sqrt{s}$ = 40 GeV.
\item[Fig.~4:]
        Single helicity asymmetry (fig.~4a) and cross section (fig.~4b)
        for J/$\psi$ production at p$_t^2$~=~5~(GeV/c)$^2$ , as
        predicted from twist--3 pQCD for $\sqrt{s}$ = 40 GeV. Both are
        shown in dependence on the J/$\psi$ energy.
\item[Fig.~5:]
        Single transverse spin asymmetry for inclusive $\pi^0$ (fig. 5a) and
        $\pi^+$ production vs.
        p$_{\perp}$, as predicted by a constituent quark model for beam
energies
        of 70, 200 and 800 GeV.
\item[Fig.~6:]
        Single transverse spin asymmetry vs. $x_F$ for inclusive pion
        production in scattering polarized (anti)protons on unpolarized
        protons at 200 GeV beam energy. The predictions shown are from the
        'Berliner' model, data points are from E704.
\item[Fig.~7:]
        Same for dilepton production in scattering unpolarized protons
        on polarized protons, deuterons, and neutrons at 800 GeV beam energy.
\item[Fig.~8:]
        Prediction of the Instanton model for the total quark contribution to
        the proton spin as a function of $Q^2$.
\item[Fig.~9:]
        Instanton model predictions for the spin--dependent structure functions
        of the proton, $g_1^p(x)$ (fig.~9a) and of the neutron, $g_1^n(x)$
        (fig.~9b).
\item[Fig.10:]
        Single spin asymmetry vs. $p_T^2$, predicted in diffractive high $p_t$
        2--jet production in case of the existence of a spin--dependent
        quark--pomeron vertex.
\end{itemize}


\end{document}